\documentclass[letterpaper,10pt]{article} 

\usepackage{opticameet3} 

\newcommand\authormark[1]{\textsuperscript{#1}}

\usepackage{amsmath,amssymb}
\usepackage[colorlinks=true,bookmarks=false,citecolor=blue,urlcolor=blue]{hyperref} 

\begin{document}

\title{Chlorophyll Absorption Analysis Enabled by Silicon-Rich Nitride-Based Concentric Ring Metalens}


\author{Alireza Khalilian\authormark{1}, Bowen Yu \authormark{1},  Mehdi Sh. Yeganeh\authormark{3}, and Yasha Yi\authormark{1,2,*}}

\address{\authormark{1} Integrated Nano Optoelectronics Laboratory, University of Michigan, 4901 Evergreen Rd., Dearborn, Michigan 48128, USA\\
\authormark{2}Energy Institute, University of Michigan, 2301 Bonisteel Blvd., Ann Arbor, Michigan 48109-2100, USA

\authormark{3}Department of Mechanical Engineering, University of Michigan, 4901 Evergreen Road, Dearborn, MI 48128, USA}

\email{\authormark{*}yashayi@umich.edu} 

\begin{abstract}A silicon-rich nitride (SRN) concentric-ring metalens, operating at 685 nm, achieves measured numerical aperture of 0.5 and 36\% focusing efficiency, enabling metalens-based chlorophyll absorption analysis and plant health evaluation for the first time.
\end{abstract}

\section{Introduction}

Silicon-rich nitride (SRN) is a practical material for metalens applications due to its tunable optical properties, such as refractive index and absorption coefficient, which can be adjusted by varying the silicon-to-nitrogen ratio \cite{ye2019silicon}. Unlike materials like titanium dioxide (TiO$_2$) and gallium nitride (GaN), which often require more complex techniques such as atomic layer deposition (ALD) or metal-organic chemical vapor deposition (MOCVD) to achieve similar tunability \cite{chen2018broadband, chen2021high}, SRN can be deposited using simpler methods like plasma-enhanced chemical vapor deposition (PECVD). Additionally, SRN is compatible with complementary metal-oxide-semiconductor (CMOS) fabrication processes, making it a viable choice for applications requiring integration with existing semiconductor technologies. Herein, we fabricated a metalens with diameter of 40 $\mu$m and experimentally measured its numerical aperture (NA) of 0.5, and a focal length of 32.5 $\mu$m, optimized for operation at 685 nm. The metalens uses a concentric ring design to achieve phase control and light focusing. The chosen operational wavelength aligns with the absorption peak of chlorophyll, making it suitable for imaging applications in plant health assessment \cite{cinque2000absorption}. Chlorophyll content imaging in leaf samples is explored to demonstrate the performance of the proposed SRN-based metalens. These findings suggest potential applications of SRN-based metalenses and SRN tunability in biological and fluorescence imaging.

\section{Lens Design and Experimental Results}

\begin{figure}[htbp]
  \centering
  \includegraphics[width=14.5cm]{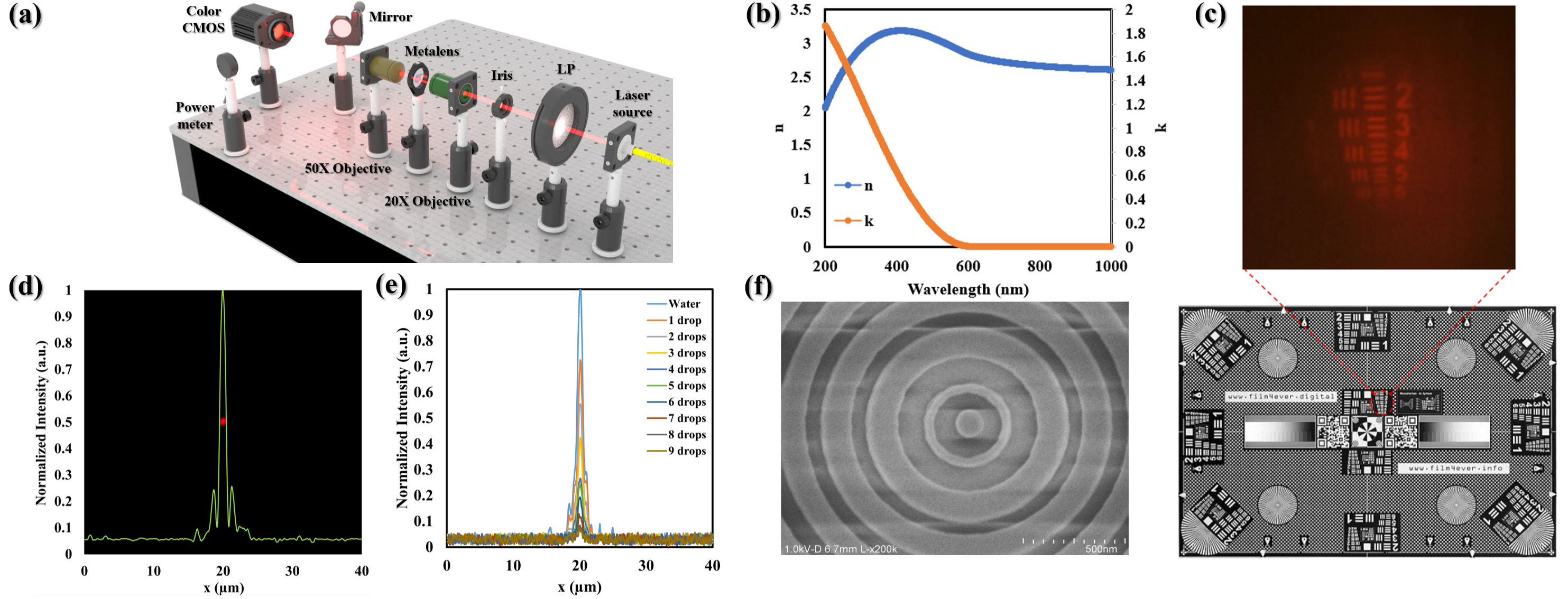}
\caption{(a) Metalens characterization setup. (b) SRN film n, k values from Woollam spectroscopy. (c) High resolution 35 mm test target imaging result. (d) Focal point under 685 nm and. (e) Focal point intensity vs chlorophyll concentration variations. (f) Top-view SEM image of a portion of the fabricated metalens. Scale bar, 500 nm.}
\label{fig:fig1}
\end{figure}

The metalens design relies on precise phase shift engineering using nanostructures. To achieve the desired phase profile through propagation phase, the lens plane is divided into discrete pitches. Each pitch is filled with high refractive index SRN (\( n = 2.74 \) at \( \lambda = 685 \, \mathrm{nm} \)) (Fig. \ref{fig:fig1} (b)), tailored to achieve variations in the effective refractive index across the lens plane. The objective is to achieve complete $2\pi$ phase coverage while maintaining high transmission efficiency. Using high refractive index SRN enables a full $2\pi$ phase shift with a smaller filling ratio, reducing pitch size and contributing to a more continuous phase distribution, which ultimately enhances the focusing efficiency of the metalens. The pitch size is set to \( 220 \, \mathrm{nm} \), allowing for near-continuous phase coverage within the fabrication-limited aspect ratio of 10:1. The smaller pitch size reduces diffraction and scattering, thereby enhancing focusing efficiency. The fabrication includes deposition of a \( 600 \, \mathrm{nm} \) thick \( \mathrm{SRN} \) layer on a glass substrate, followed by a \( 300 \, \mathrm{nm} \) thick \( \mathrm{SiO_2} \) hard mask. A \( 200 \, \mathrm{nm} \) thick ZEP520A e-beam resist is spin-coated onto the hard mask. The metalens pattern is written using electron beam lithography (e-beam) and transferred into the \( \mathrm{SiO_2} \) layer via reactive ion etching (RIE). A second RIE step transfers the pattern into the \( \mathrm{SRN} \) layer. This two-step etching process compensates for the limited selectivity of e-beam resists against \( \mathrm{SRN} \). Characterization of the metalens was performed using the setup in Fig. \ref{fig:fig1} (a). Figure \ref{fig:fig1} (c) Depicts the lens ability in realizing high resolution target.The lens demonstrated focusing efficiency of 36\% with full width at half maximum (FWHM) of $1.42\lambda$ (Fig. \ref{fig:fig1} (d)), and an \( \mathrm{NA} \) of 0.5. Figure \ref{fig:fig1} (e) indicating the focal point intensity variations with different concentration of chlorophyll content in 70 mL of water (each drop corresponds to 1.6 mg of chlorophyll). Figure \ref{fig:fig1} (f) represents a top-view SEM image of a portion of the fabricated metalens with scale bar, 500 nm.

\begin{figure}[htbp]
  \centering
  \includegraphics[width=15cm]{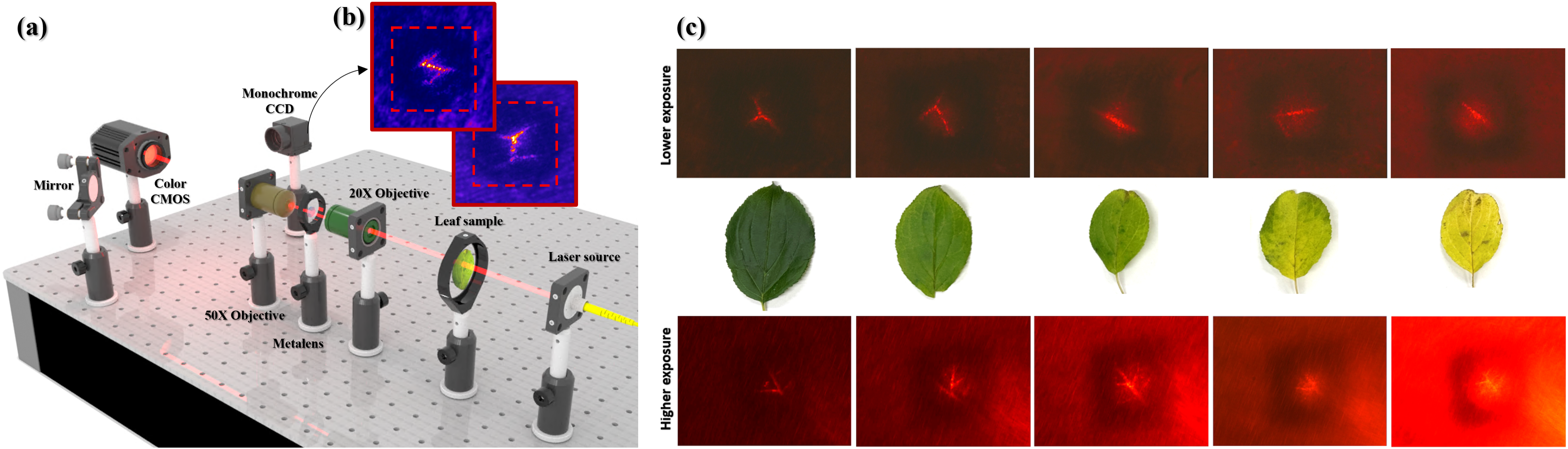}
\caption{(a) Leaf imaging setup. Imaging results of leaf samples (b) using monochrome CCD with pseudocolored fire effect and (d) using color CMOS sensors with varying chlorophyll content.}
\label{fig:fig2}
\end{figure}

To demonstrate the application of our fabricated metalens, we performed imaging of leaf samples using the setup shown in Fig. \ref{fig:fig2} (a). A monochrome CCD used to capture a healthy leaf images in Fig. \ref{fig:fig2} (b). To investigate this capability more, leaf samples with high to low chlorophyll were used next. Each column in Fig. \ref{fig:fig2} (c) represents a single leaf sample. The images in the middle row, captured with a phone camera, provides a visual reference for the leaf’s condition, while the top and bottom rows show images captured by color CMOS at low and high exposure settings, respectively. Darker regions in these images correspond to higher chlorophyll content, particularly noticeable in the healthier samples on the left. In contrast, weaker absorption is observed in the samples on the right, indicating reduced chlorophyll levels associated with stress or senescence. Metalens flatness, light weight, and compatibility with semiconductor manufacture technology, it is offering advantages over conventional systems that typically require additional bulky conventional lenses. To the best of our knowledge, this study presents the first application of a metalens in plant health assessment. Future studies will explore the potential of SRN-based metalenses in similar contexts, leveraging their optical tunability and filtering effects to advance chlorophyll imaging and plant biology studies.

\bibliographystyle{opticajnl}
\bibliography{sample} 

\end{document}